\documentclass[twocolumn,showpacs,preprintnumbers,amsmath,amssymb]{revtex4}

\def\undersim#1{\setbox9\hbox{${#1}$}{#1}\kern-\wd9\lower
    2.5pt \hbox{\lower\dp9\hbox to \wd9{\hss $_\sim$\hss}}}
\usepackage{amssymb}
\usepackage{amsmath}
\usepackage{graphicx}
\usepackage{color}

\def\undersim#1{\setbox9\hbox{${#1}$}{#1}\kern-\wd9\lower
    2.5pt \hbox{\lower\dp9\hbox to \wd9{\hss $_\sim$\hss}}}

\def\mk{{\mathbf k}}

\begin{document}

\title{Hanbury Brown-Twiss correlation of $\phi\phi$ with in-medium mass modification}

\author{Yong Zhang$^{1}$}
\author{Peng Ru$^{2}$}
\affiliation{\small$^1$School of Mathematics and Physics,
Jiangsu University of Technology, Changzhou, Jiangsu 213001, China\\
$^2$School of Materials and New Energy, South China Normal University, Shanwei 516699, China}

%\date{\today}

\begin{abstract}

We study the impacts of the squeezing effect caused by in-medium mass modification on the Hanbury Brown-Twiss (HBT) correlation of $\phi\phi$
using a hydrodynamical source.
The squeezing effect reduces the influence of transverse flow on the transverse distribution of $\phi$ emitting source while expanding the longitudinal source distribution. This trend becomes more noticeable as the transverse momentum increases, resulting in an enlargement of the HBT radii for $\phi\phi$.
Notably, this enlargement is particularly evident with higher transverse pair momentum.
As a result, the HBT radii of $\phi\phi$ increase as transverse pair momentum increases instead of showing a consistent decrease,
showcasing non-flow behavior of HBT radii in hydrodynamical sources.
Both Gaussian fitting and L\'{e}vy fitting support the conclusion.

\end{abstract}
\maketitle

\section{Introduction}
In high-energy heavy-ion collisions, people investigate the properties of the resulting matter and system by
analyzing particle observables in the final state.
Hanbury Brown-Twiss (HBT) correlation is a significant measurement in these collisions,
offering insights into the spatial and temporal structure of particle emissions \cite{Gyu79,Wongbook,Wie99,Wei00,Csorgo02,Lisa05}.

Before particles are detected, their interactions with the source medium may lead to a modification in mass.
These changes could lead to a squeezing effect that brings about an interesting experimental phenomenon called squeezed back-to-back correlation (SBBC) between bosons
and antibosons \cite{AsaCso96,AsaCsoGyu99,Padula06,DudPad10,Zhang15a,Zhang-EPJC16,AGY17,XuZhang19,Zhang2024}.
This squeezing effect is connected to the boson's in-medium mass modification via a Bogoliubov transformation. Such transformation forms a link between the creation (or annihilation) operators of quasiparticles in the medium and their corresponding particles in a vacuum \cite{AsaCso96,AsaCsoGyu99,Padula06,DudPad10,Zhang15a,Zhang-EPJC16,AGY17,XuZhang19,Zhang2024}.
The squeezing effect also impacts the HBT correlation \cite{DudPad10,Zhang2024}. The squeezing {\color{black}effect}
diminishes the impact of flow on the HBT, resulting in a non-flow behavior in the HBT radii \cite{Zhang2024}.
Given that the SBBC signal may fully disappear due to the source's wide temporal distribution \cite{Padula06,DudPad10,Zhang15a,Zhang-EPJC16},
the non-flow behavior of the HBT radii still exists \cite{Zhang2024}.
Consequently, beyond SBBC, the non-flow behavior of the HBT radii could present a new way to study the squeezing effects.

The previous conclusions about the impact of the squeezing effect on HBT were based on an expanding source with spatiotemporal independence \cite{DudPad10,Zhang2024}.
Studying the impact of the squeezing effect on HBT through more realistic space-time evolving source models is needed.
In this paper, the ideal relativistic hydrodynamics in $2 + 1$ dimensions is used to model the transverse expansion of sources,
while the source longitudinal evolution is described by the Bjorken boost-invariant hypothesis \cite{Bjorken_PRD83}.
To begin exploring the impact of squeezing on HBT with a more realistic model,
these descriptions are appropriate for the heavy-ion collisions at the RHIC top energy and the LHC energy \cite{{Ris98,KolHei03,Bay83,Gyu97,Ris9596,BLM96,
BLM04,Kol00,KolRap03,She10,HuichaoS}}.
Furthermore, the simulation calculation of HBT correlation functions with squeezing
effect has now evolved to incorporate randomness in the direction of the momenta of two bosons,
unlike the previous method which always assumed the same directions for both bosons \cite{Zhang2024}.

The squeezing effect on HBT correlation of $\phi$$\phi$ are studied in this paper.
The squeezing effect not only suppresses the influence of transverse flow on the transverse source
distribution but also widens the longitudinal source distribution, further impacting the HBT {\color{black}correlation function and HBT radii} of $\phi\phi$.
The impacts of the squeezing effect on curve {\color{black} of the correlation function} and HBT radii are more pronounced for $\phi$$\phi$ with large transverse pair momentum,
resulting in a non-monotonic decrease of HBT radii as $K_T$ increases. This phenomenon is called non-flow behavior of the HBT radii \cite{Zhang2024}.
Both Gaussian fitting and L\'{e}vy fitting support the above conclusion. 

$\phi$ meson is an excellent probe for studying the quark-gluon plasma (QGP) formed in high-energy heavy-ion collisions
because it contains a strange quark that is thought to undergo the full evolution of the QGP created in these collisions.
Significant interest has been generated by the recent analyses of experimental data concerning the $\phi$ meson \cite{ALICE-EPJC18p,STAR-PRL16p,
STAR-PRC16p,PHENIX-PRC16p,ALICE-PRC15p,PHENIX-PRC11p,STAR-PLB09p,STAR-PRC09p,STAR-PRL07p,PHENIX-PRL07p,PHENIX-PRC05p,NA50-PLB03p,STAR-PRC02p,NA50-PLB00p}.
The modification of the $\phi$ meson mass in the medium was expected to be present within the particle-emitting sources generated in high-energy heavy-ion collisions \cite{pm,pm1,pm2}.
Consequently, the study of this paper is meaningful in such collisions.

The organization of this paper is as follows: Section II gives the formulas of the HBT correlation function with a squeezing effect for a hydrodynamical source.
Section III demonstrates the impacts of the squeezing effect on the spatial and temporal distribution of the $\phi$ emission, and the influences of the squeezing effect on
HBT {\color{black}correlation function and HBT radii} are also analyzed in this section.
Concluding the paper, Section IV offers a summary and discussion.

\section{Formulas}
The HBT correlation function of two identical bosons was defined as \cite{Gyu79,Wongbook,Wie99,Wei00,Csorgo02,Lisa05}

\begin{eqnarray}
C(\mk_1,\mk_2) =1+\frac{|\int d^{4}rS(r,K)e^{iq\cdot r}|^{2}}{\int d^{4}r_{1}S(r_{1},k_1)d^{4}r_2S(r_2,k_2)}.
\end{eqnarray}
Here, $S(r,k)$ is the emission function. The momenta and positions of the two bosons are represented by
$k_1$, $r_1$ and $k_2$, $r_2$ respectively. Furthermore, $q = k_1-k_2$, $K = (k_1+k_2)/2$ stand for the relative and pair momentum of two identical bosons.
For hydrodynamic sources, the emission function can be expressed as \cite{Cooper1974,Sch1992,Heinz-PLB1994,Heinz-PRC2015,Heinz-PRC2018}
\begin{equation}\label{emission}
S(r,k) = \frac{1}{(2\pi)^3}\int_{\Sigma}k^{\mu}  d^3 \sigma_{\mu}(r^{\prime})\, \delta^4 (r-r^{\prime})f(r,k),
\end{equation}
when the squeezing effect is not considered, the $f(r,k)$ is
\begin{equation}\label{f0}
f(r,k) = \frac{1}{\exp[k^{\mu}u_{\mu}(r^{\prime})/T]-1}.
\end{equation}
Here, the flow velocity profiles along the freeze-out surface $\Sigma$ is denoted by $u_{\mu}(r)$, while the outward pointing normal vector on
$\Sigma$ at point $r$ is represented by $d^3\sigma_{\mu}(r)$. $T$ is the freeze-out temperature of particle.
When the squeezing effect is considered, the $f(r,k)$ becomes \cite{Gyu79,Wongbook,Wie99,Wei00,Csorgo02,Lisa05,AsaCso96,AsaCsoGyu99,Padula06,DudPad10,Zhang2024}

\begin{equation}\label{f}
f(r,k) = |c_{\mk'}|^2\,n_{\mk'}+\,|s_{-\mk'}|^2\,(\,n_{-\mk'}+1),
\end{equation}
\begin{equation}\label{csk}
c_{\mk'}=\cosh[\,r_{\mk'}\,], \,\,\,s_{\mk'}=\sinh[\,r_{\mk'}\,],
\end{equation}
\begin{eqnarray}
r_{\mk'}=\frac{1}{2} \log \left[\omega_{\mk'}/\Omega_{\mk'}\right],
\end{eqnarray}
\begin{eqnarray}
\omega_{\mk'}(r)=\sqrt{\mk'^2(r)+m^2}=k^{\mu} u_{\mu}(r),
\end{eqnarray}
\begin{eqnarray}
&&\hspace*{-7mm}\Omega_{\mk'}(r)=\sqrt{\mk'^2(r)+m_*^2}\nonumber\\
&&\hspace*{3.9mm}=\sqrt{[\omega_{\mk'}(r)]^2-m^2+m_*^2},
\end{eqnarray}
\begin{equation}\label{BZ}
n_{\mk'}=\frac{1}{\exp(\Omega_{\mk'}(r)/T)-1}.
\end{equation}
Here, the four-momentum of the particle is denoted by $k^{\mu}=(\omega_{\mk}=\sqrt{\mk^2+m^2},{\mk})$, while $\mk'$
represents the local-frame momentum corresponding to $\mk$. $m$ represents the mass of the particle in a vacuum, while $m_*$ is
the mass of the particle in the source medium. When $m_* =$ $ m$, it indicates that there is no mass modification in the medium, and
$f(r,k)$ in Eq. \ref{f} becomes equal to the $f(r,k)$ in Eq. \ref{f0}.
It is worth mentioning that the in-medium mass modification of bosons is considered in this paper, and the change of width in the medium is not considered.

In this paper's calculations, a Gaussian distribution is used as the initial energy density distribution in the transverse plane at
$\tau_0=0.6$ fm/$c$ \cite{Zhang15a}:
\begin{eqnarray}
\epsilon = \epsilon_0\exp[-x^2/(2R_x^2)-y^2/(2R_y^2)],
\end{eqnarray}
$\epsilon_0$ represents the initial energy density at the center of the transverse plane, while $R_x$ and $R_y$ represent
the radii of the energy density distribution in the $x$ and $y$ direction in the transverse plane.

\section{Results}
In this section, the impacts of the squeezing effect on the spatial and temporal distribution of the $\phi$ emission source are first discussed, and then
the impacts of the squeezing effect on HBT {\color{black}correlation function and HBT radii} of $\phi$$\phi$ are shown. In the calculations, the space-time rapidity is considered within the range of $(-1,1)$.
The $\phi$ meson's freeze-out temperature is considered to be 0.14 GeV \cite{Padula06,Zhang2024}.
The mass of $\phi$ meson in a vacuum is denoted by $m$ and is considered to be 1.01946 GeV \cite{PDG24}.
The in-medium mass modification is denoted as $\delta m$, and $\delta m = m-m_* $.
When $\delta m = 0 $, it means that the squeezing effect is not taken into account in the calculations.
In the medium, it is anticipated that the mass of $\phi$ meson will be decreased by approximately 0.01 to 0.02 GeV \cite{pm}.
Hence, $\delta m$ is taken as two values, 0.01 and 0.02 GeV.

\subsection{Spatial and temporal distribution of the emission source}
\begin{figure}[htbp]
\vspace*{3mm}
\includegraphics[scale=0.59]{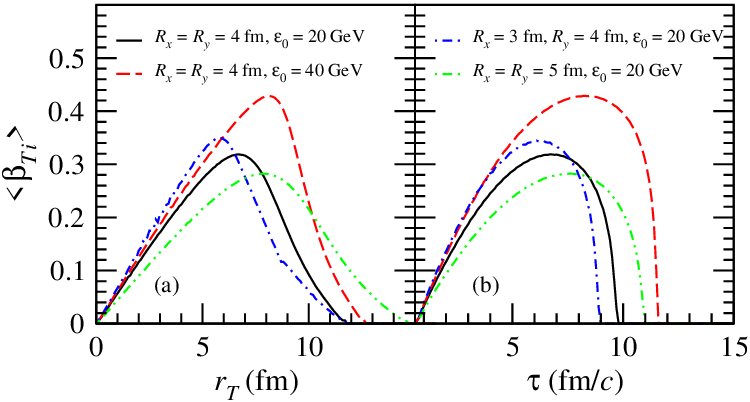}
\vspace*{0mm}
\caption{(Color online) The average transverse flow $\langle \beta _{Ti} \rangle$ of the $\phi$ emission sources. Plot (a) is the results as a function of transverse coordinate
$r_T$, while plot (b) shows the results as a function of proper time $\tau$.  }
\label{vt}
\end{figure}

The average transverse flow $\langle \beta _{Ti} \rangle$ of the $\phi$ emission sources is shown in Fig. \ref{vt}.
Here, $\beta _{Ti} =\sqrt{v_{xi}^2+v_{yi}^2}$ is the transverse expanding velocity at the $i$ freeze-out point.
$r_T = \sqrt{x^2+y^2}$ and $\tau$ are the transverse coordinate and proper time of the $\phi$ meson, respectively.
Four sets of initial conditions are used in this paper.
The initial conditions for $R_x = R_y$ may approximate to central collisions,
while those for $R_x < R_y$ could correspond to non-central collisions.
A value of $\epsilon_0 = 20$ GeV is close to the center initial energy density of the Au$+$Au collisions at $\sqrt{s_{NN}}=200$ GeV,
whereas $\epsilon_0 = 40$ GeV is close to the center initial energy density of the Pb$+$Pb collisions at $\sqrt{s_{NN}}=2.76$ TeV \cite{YH}.
For fixed $R_x$ and $R_y$, the transverse flow increases as the center initial energy density $\epsilon_0$ increases.
For fixed $\epsilon_0$, the transverse flow decreases as the $R_x$ and $R_y$ increase.

\begin{figure}[htbp]
\vspace*{0mm}
\includegraphics[scale=0.59]{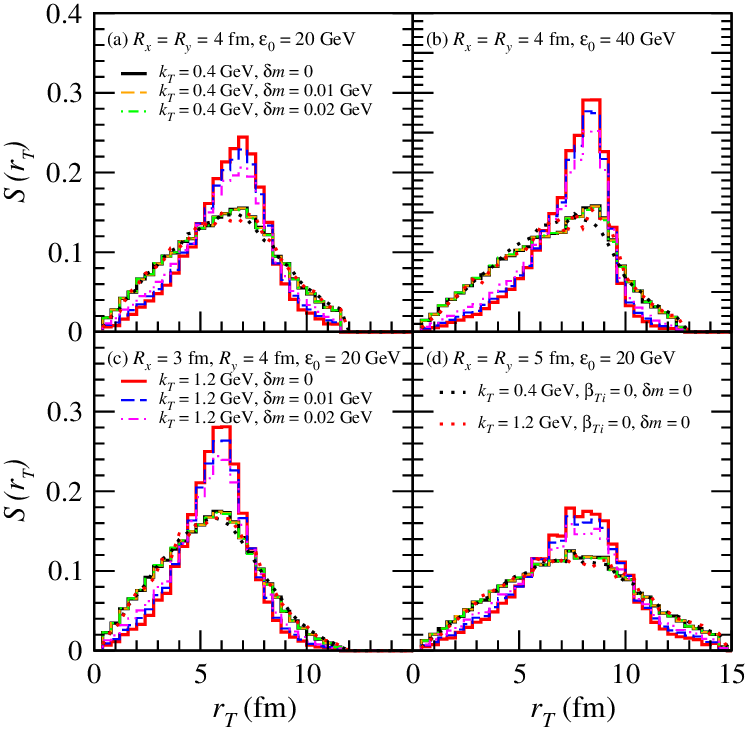}
\vspace*{0mm}
\caption{(Color online) Normalized distributions of transverse coordinate of the $\phi$ emission sources for $k_T$ = 0.4 GeV and 1.2 GeV. }
\label{srt}
\end{figure}

In Fig. \ref{srt}, the normalized distributions of transverse coordinate of the $\phi$ emission sources for $k_T$ = 0.4 GeV and 1.2 GeV are shown.
Here, $k_T = \sqrt{k_x^2+k_y^2}$ is the transverse momentum.
The black and red dot lines represent the results for transverse flow $\beta _{Ti} = 0$ (These results are obtained by artificially setting the transverse flow velocity at $\beta _{Ti} = 0$ at the freezing point during the calculation.), while the transverse flow at the freezing point of the other lines in Fig. \ref{srt} is provided by relativistic hydrodynamics. For $\beta _{Ti} = 0$, the transverse distribution of the sources with the same initial condition is nearly identical for $k_T$ values of 0.4 GeV and 1.2 GeV.
The results suggest that in the absence of squeezing effect ($\delta m = 0 $),
the transverse flow has little effect on the distribution of the $\phi$ emission sources with $k_T$ = 0.4 GeV.
However, it tends to narrow the transverse distribution of the sources with $k_T$ = 1.2 GeV.
This phenomenon becomes more significant as the transverse flow velocity increases.
The reason for this is that the transverse flow results in an increased generation of particles with high transverse
momentum in regions with high transverse flow.
When $\delta m > 0 $, the squeezing effect is taken into account, showing minimal influence on the distribution of emission sources with $k_T$ = 0.4 GeV.
However, the squeezing effect diminishes the impact of transverse flow on the distribution of sources at $k_T$ = 1.2 GeV,
resulting in a broadened transverse distribution compared to when $\delta m = 0 $.
This phenomenon becomes more obvious as the $\delta m$ increases.

\begin{figure}[htbp]
\vspace*{0mm}
\includegraphics[scale=0.59]{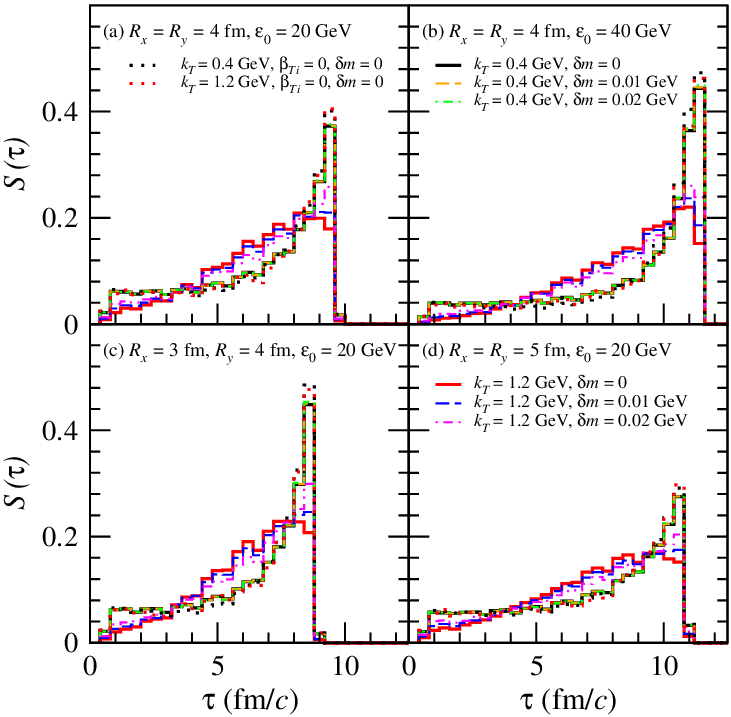}
\vspace*{0mm}
\caption{(Color online) Normalized distributions of proper time $\tau$ of the $\phi$ emission sources for $k_T$ = 0.4 GeV and 1.2 GeV. }
\label{sta}
\end{figure}

In Fig. \ref{sta}, the normalized distributions of proper time $\tau$ of the $\phi$ emission sources for $k_T$ = 0.4 GeV and 1.2 GeV are shown.
For $\beta _{Ti} = 0$, the source distributions of the proper time $\tau$ is almost identical for $k_T$ values of 0.4 GeV and 1.2 GeV.
When the squeezing effect is not present ($\delta m = 0 $), the transverse flow minimally influences the source distributions of proper time at $k_T$ = 0.4 GeV.
However, at $k_T$ = 1.2 GeV, it broadens the distributions of proper time. By comparing the results of transverse flow in Fig. \ref{vt}(b),
it can be concluded that the reason for the above phenomenon is the transverse flow results in an increased generation of particles with high transverse
momentum in regions with high transverse flow.
The squeezing effect ($\delta m > 0 $) has little influence on the source distribution of $\tau$ with $k_T$ = 0.4 GeV.
Conversely, the squeezing effect suppresses the impact of transverse flow on the source distribution of $\tau$ at $k_T$ = 1.2 GeV,
particularly notable for higher values of $\delta m$.

\begin{figure}[htbp]
\vspace*{3mm}
\includegraphics[scale=0.59]{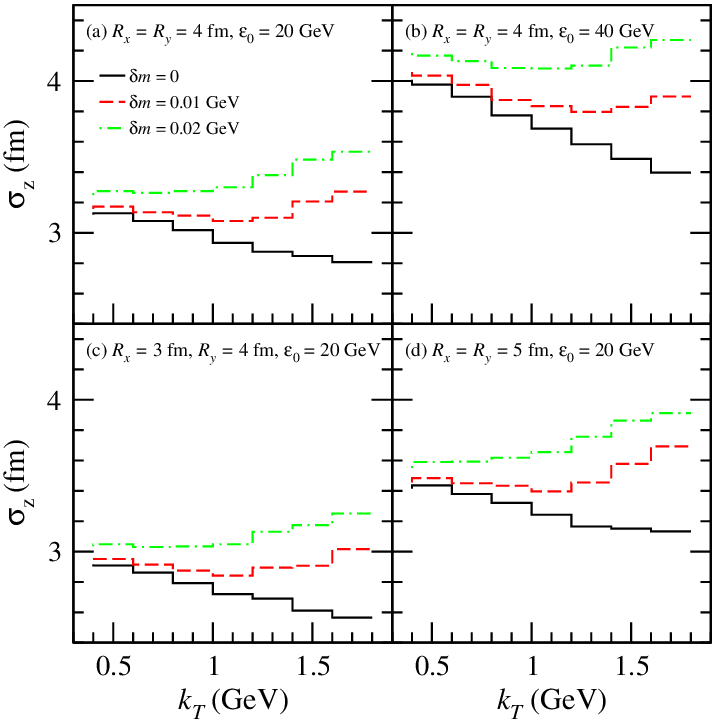}
\vspace*{0mm}
\caption{(Color online) The root mean square radius ($\sigma _z$) in longitudinal direction. }
\label{sz}
\end{figure}

In Fig. \ref{sz}, the root mean square radius $\sigma _z$ in longitudinal direction is shown as a function of $k_T$.
Here, $\sigma_{z} = \sqrt{\frac{1}{N}\sum_{i=1}^{N}(\,{z}_{i}-\bar{{z}}\,)^{\,2}},$ $z_i$ represents the longitudinal coordinate of a particle marked as $i$,
with its average value represented by $\bar{z_i}$, and $N$ is the number of particles in each $k_T$ bin.
The width of the source distribution in the longitudinal direction might be qualitatively described by $\sigma _z$.
When squeezing effect is not considered ($\delta m = 0 $),
the width of the source distribution in the longitudinal direction
decreases as transverse momentum $k_T$ increases.
Conversely, accounting for the squeezing effect broadens the source distribution in the longitudinal direction, particularly noticeable with higher $k_T$,
leading to an increase in its width as $k_T$ increases.
The impacts of the squeezing effect on the source distribution in the longitudinal direction increases as the $\delta m$ increases.
For fixed $k_T$ and $\delta m$, the squeezing effect expands the longitudinal distribution width of the source more than the transverse distribution width.

\subsection{Squeezing effect on HBT}
\begin{figure}[htbp]
\vspace*{2mm}
\includegraphics[scale=0.59]{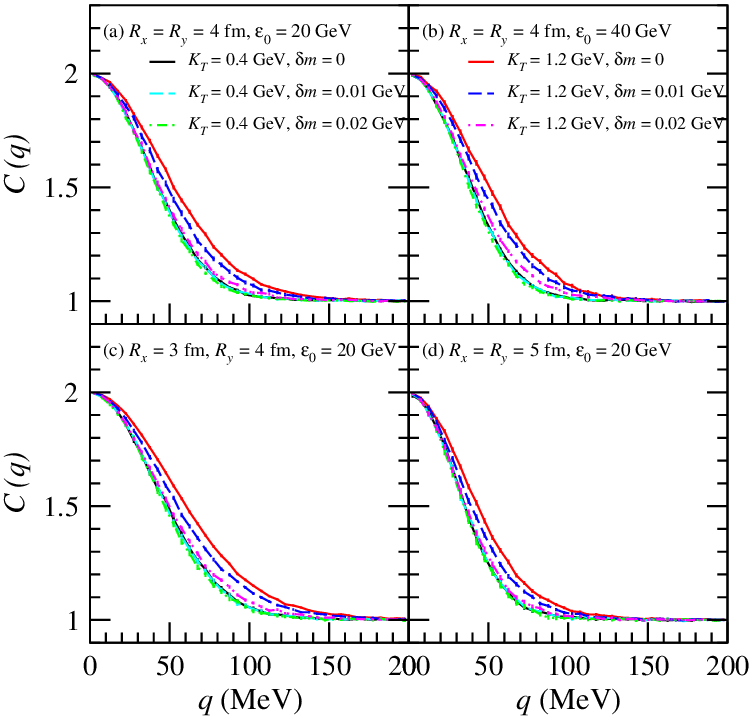}
\vspace*{0mm}
\caption{(Color online) $\phi\phi$ {\color{black}correlation functions} with respect to relative momentum $q$ for $K_T$ = 0.4 GeV and 1.2 GeV. Here, $K_{T}$ is the transverse momentum of $\phi$ pair and $2K_{T}=|\mk_{T1}+\mk_{T2}|$. }
\label{hbtr}
\end{figure}

In Fig. \ref{hbtr}, {\color{black}the $\phi\phi$ correlation functions} with respect to relative momentum $q$ are shown.
Here, $K_{T}$ is the transverse momentum of $\phi$ pair and $2K_{T}=|\mk_{T1}+\mk_{T2}|$.
When the squeezing effect is not considered, the curve {\color{black} of the  correlation function}
for $K_T$ = 1.2 GeV appears broader than that for $K_T$ = 0.4 GeV.
Two factors contribute to this phenomenon: firstly, the transverse distribution width of $k_T$ = 1.2 GeV
due to transverse flow is smaller compared to that of $k_T$ = 0.4 GeV (see Fig. \ref{srt});
secondly, the longitudinal distribution width of the sources decreases as $k_T$ increases (see Fig. \ref{sz}).
The squeezing effect narrows the curve {\color{black} of the  correlation function} and is more obvious for large $K_T$. This is due to
the squeezing effect suppresses the impact of transverse flow on the transverse source distribution
and widens the longitudinal source distribution.
The impacts of the squeezing effect on the curve {\color{black} of the  correlation function}
increases as $\delta m$ increases.

To further study quantitatively the impact of the squeezing effect on HBT, the one-dimensional HBT radii $R$
is obtained by fitting the HBT correlation function.
{\color{black}In fitting, the parametrized formula is selected as}

\begin{eqnarray}
\label{levnh}
C(q)=1+\lambda e^{-|qR|^\alpha}.
\end{eqnarray}
{\color{black}$\lambda$ is the chaoticity parameter.}
$\alpha$ is commonly known as the L\'{e}vy exponent. It also represents the shape of the source, with $\alpha = 2$ leading to the Gaussian distribution and $\alpha = 1$ resulting in the Cauchy distribution {\color{black}\cite{Tlevy,levy1,levy2,levy3,levy4}}.
{\color{black}This paper adopts two fitting methods. The first method, referred to as Gaussian fitting, involves utilizing a constant value of 2 for $\alpha$.
In contrast, the second method, known as L\'{e}vy fitting, utilizes $\alpha$ as a fitting parameter \cite{Tlevy,levy1,levy2,levy3,levy4}.}
Since the $\phi$ meson is electrically neutral, it is unaffected by the Coulomb effect. Thus, $\phi$ meson is a good probe for studying the squeezing effect in contrast to charged bosons. When conducting HBT radii fitting for the $\phi$ meson, there is no necessity to consider the Coulomb effect.

\begin{figure}[htbp]
\vspace*{3mm}
\includegraphics[scale=0.59]{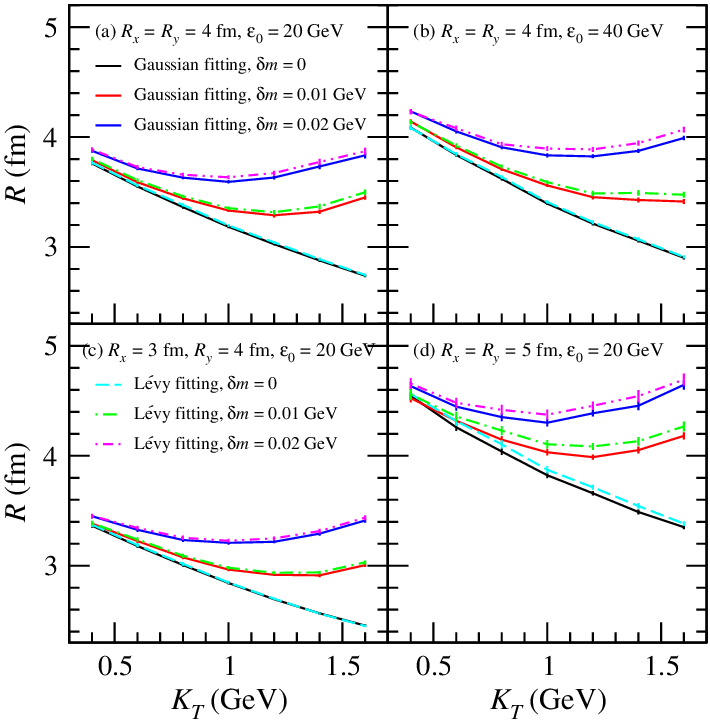}
\vspace*{0mm}
\caption{(Color online) HBT radii of $\phi\phi$ with respect to transverse pair momentum $K_T$. }
\label{rgsl}
\end{figure}

In Fig. \ref{rgsl}, the HBT radii of $\phi\phi$ with respect to transverse pair momentum $K_T$ are shown.
The HBT radii obtained by L\'{e}vy fitting are slightly larger than those obtained by Gaussian fitting,
primarily because the L\'{e}vy exponent $\alpha$ is slightly less than 2.
If the squeezing effect is not considered, the HBT radii decreases monotonically as $K_T$ increases.
However, when accounting for the squeezing effect, the HBT radii increase, particularly for higher $K_T$.
Consequently, this causes the HBT radii to no longer decreasing monotonically with the increase of $K_T$,
and even increasing with the increase of $K_T$.
The impacts of the squeezing effect on the HBT radii
increases as $\delta m$ increases.
The phenomenon where the HBT radii do not monotonically decrease with increasing $K_T$ is known as the non-flow behavior of the HBT radii \cite{Zhang2024}.
{\color{black}It should be mentioned that the effects of energy-momentum conservation are not considered in the calculations. Since the initial conditions used in this paper are close to high multiplicity heavy-ion collisions, the effects of energy-momentum conservation on two-particle correlations at low relative momentum are negligible \cite{Wie99,Lisa05}.
In the small transverse pair momentum range, the influence of the squeezing effect on both the correlation function and HBT radii is also very weak. Energy-momentum conservation might interfere with the impact of the squeezing effect on the correlation function and HBT radii.
Conversely, at higher transverse pair momentum levels, the squeezing effect significantly affects the correlation function and HBT radii, and energy-momentum conservation may not interfere with its observation. Moreover, the non-flow behavior of the HBT radii mainly occurs at higher transverse pair momentum ranges,
where energy-momentum conservation may not interfere with the observation of this behavior.}

\begin{figure}[htbp]
\vspace*{3mm}
\includegraphics[scale=0.59]{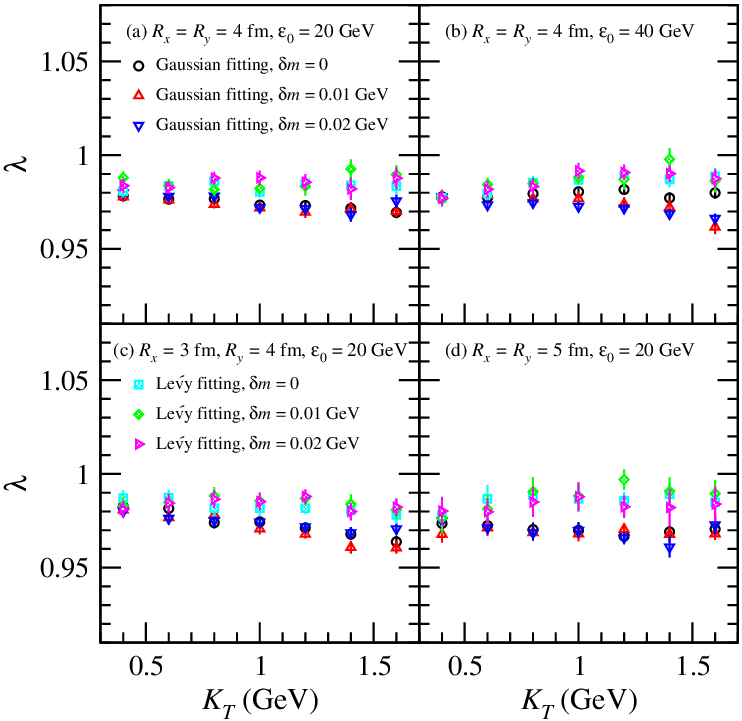}
\vspace*{0mm}
\caption{(Color online) Chaoticity parameter $\lambda$ with respect to transverse pair momentum $K_T$. }
\label{lambda}
\end{figure}
{\color{black}In Fig. \ref{lambda}, the chaoticity parameter $\lambda$ with respect to transverse pair momentum $K_T$ are shown.
In most cases, $\lambda$ is slightly less than 1 for various values of $K_T$, with only a few instances where it is nearly equal to 1.
For Gaussian fitting and the L\'{e}vy fitting, $\lambda$ is almost unaffected by the squeezing effect.}

\begin{figure}[htbp]
\vspace*{3mm}
\includegraphics[scale=0.59]{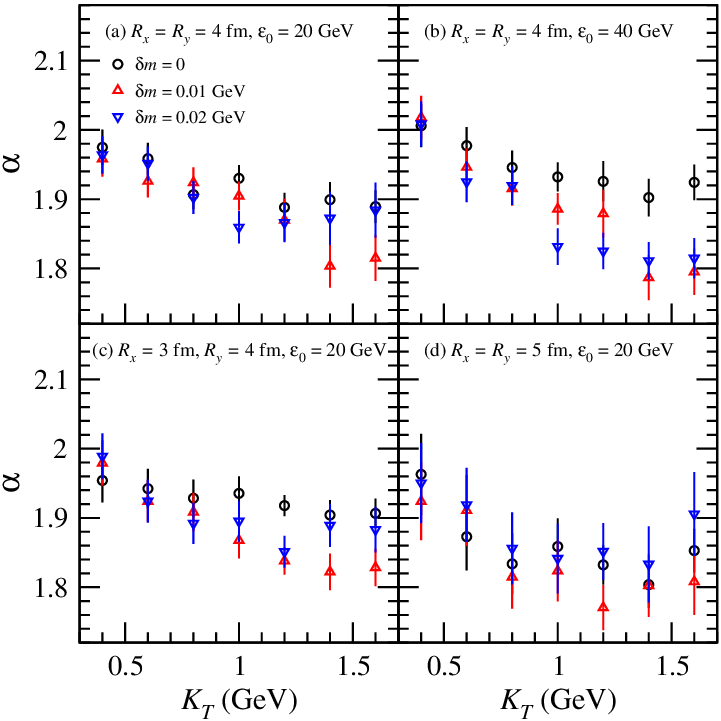}
\vspace*{0mm}
\caption{(Color online) L\'{e}vy exponent $\alpha$ with respect to transverse pair momentum $K_T$. }
\label{leva}
\end{figure}

In Fig. \ref{leva}, the L\'{e}vy exponent $\alpha$ with respect to transverse pair momentum $K_T$ are shown.
{\color{black}A systematic tendency of decreasing $\alpha$ is observed with increasing $K_T$.
The current findings indicate that the squeezing effect has almost no impact on $\alpha$ for most cases. The actual situation may be one of the following two situations:
firstly, the squeezing effect does not impact $\alpha$; secondly, the squeezing effect impacts both transverse and longitudinal shapes and may be inconsistent, which cannot be accurately provided feedback through one-dimensional fitting.}

\section{Summary and discussion}
The particles' interactions within the source medium result in a mass modification of bosons in the medium, causing a squeezing effect.
In this paper, the impacts of the squeezing effect on the HBT correlation of $\phi\phi$ are analyzed based on a hydrodynamical source.

The squeezing effect impacts the HBT correlation by affecting the spatiotemporal distribution of the $\phi$ source.
The squeezing effect diminishes the impact of transverse flow on the transverse source distribution
and broadens the longitudinal source distribution.
This phenomenon becomes more pronounced with higher transverse momentum $k_T$, causing an increase in the HBT radii of $\phi\phi$.
Specifically, this increase stands out with greater transverse pair momentum $K_T$.
Consequently, the HBT radii increase as $K_T$ rises instead of following a consistent decrease, showcasing non-flow behavior \cite{Zhang2024}.
Both Gaussian fitting and L\'{e}vy fitting support the above conclusion.
{\color{black}The initial conditions used in this paper are close to those of the Au$+$Au collisions at $\sqrt{s_{NN}}=200$ GeV and the Pb$+$Pb collisions at $\sqrt{s_{NN}}=2.76$ TeV. Therefore, the non-flow behavior of the HBT radii of $\phi\phi$ may be detected in such collisions. Meanwhile, if the SBBC signal is not entirely suppressed due to the source's broad temporal distribution, it could also be detected. With increasing collision energy, the temporal distributions of the sources broaden.
The SBBC may be completely suppressed. However, the non-flow behavior of the HBT radii of $\phi\phi$ will not disappear \cite{Zhang2024}, and it might be detectable in collisions at extremely high energy.
Besides SBBC, the non-flow behavior of the HBT radii might provide a new avenue to explore the squeezing effects, notably for collisions at extremely high energy.
It should be mentioned that the effects demonstrated here will be present when the particles experience a change in their mass in the medium.
In the absence of in-medium mass modification, there will be no signal regarding the SBBC and the non-flow behavior of the HBT radii.}

The analysis in this paper can reflect the impacts of the squeezing effect on the spatial distribution of the $\phi$ source from the HBT radii.
In fact, the squeezing effect has different impacts on the transverse and longitudinal distribution of the sources. However, the one-dimensional HBT radii
cannot reflect the specific information of this difference.
In addition, the squeezing effect also affects the temporal distribution of the source, but the one-dimensional HBT radii cannot reflect this effect.
Thus, more detailed analysis is needed, such as three-dimensional HBT radii analysis.

In this paper's calculation, the initial energy condition is treated as Gaussian distribution. Actually, the initial energy condition of the source is fluctuating.
Studying the effects of the squeezing effect on HBT correlations by further using fluctuating initial conditions could present an interesting direction for additional exploration.

\begin{acknowledgements}
This research was supported by the National Natural Science Foundation of China
under Grant No. 11905085; Guangdong Basic and Applied Basic Research Foundation under Grant No. 2022A1515110392.
\end{acknowledgements}

\end{document}